\documentclass{kapproc} 
\usepackage{procps} 
\usepackage[dvips]{graphicx}
\upperandlowercase
\kluwerbib 

\begin{document}

\articletitle[Evolution of galaxies in clusters]
{Evolution of galaxies in clusters: \\
semi-analytic models and
  observations} 
\author{Gabriella De Lucia,\altaffilmark{1} Guinevere
  Kauffmann,\altaffilmark{1} 
  and Simon D. M. White\altaffilmark{1}}

\affil{\altaffilmark{1}Max-Planck Institute f\"ur Astrophysik, Garching,
  Germany} 
\email{gdelucia@mpa-garching.mpg.de}

\anxx{De Lucia\, Gabriella}

\begin{abstract}
We present two different projects. The first one is related to the development
of a semi-analytic model to follow the formation, the evolution and chemical
enrichment of cluster galaxies in a hierarchical dark matter model. The second
project concerns the comparison between high resolution N-body simulations of
clusters of galaxies with the EDisCS observational dataset. 
\end{abstract}

\begin{keywords}
intergalactic medium, galaxy cluster, galaxy formation, galaxy evolution
\end{keywords}

\section{The simulations}

We use collisionless simulations of clusters of galaxies generated using the
technique of "zooming-in" (Tormen et al., 1997). A suite of cluster simulations
covering a wide range of masses and structural properties has been carried out
by Barbara Lanzoni as part of her Ph.D. thesis (Lanzoni et al., 2002). 

\section{The model}

The model we use is based on the semi-analytic model presented by Springel et
al. (2001) with new prescriptions for the chemical enrichment and the transfer
of mass and metals between the different phases. The use of simple physical
prescriptions for the transport of metals in the different phases and the
choice of a few, physically motivated, parameters allow us to reproduce the
relation between stellar mass and metallicity inferred from new SDSS data as
well as the observed amount of metals in the ICM.
The good agreement between the model predictions and the observational data
shows that the circulation of metals between the different baryonic components
of the cluster is being well tracked.

Our analysis shows that the chemical pollution of the ICM occurs at relatively
high redshift: $\sim 80$ per cent of the metals today present in the ICM were
ejected at redshifts larger than $1$. Massive galaxies are important
contributors to the chemical enrichment of the ICM: $\sim 75$ per cent of the
metals today in the ICM were ejected by galaxies with mass larger than
$10^{10}\,{\rm M}_{\odot}$.  
\begin{figure}[ht]
\vskip2.7in
\includegraphics{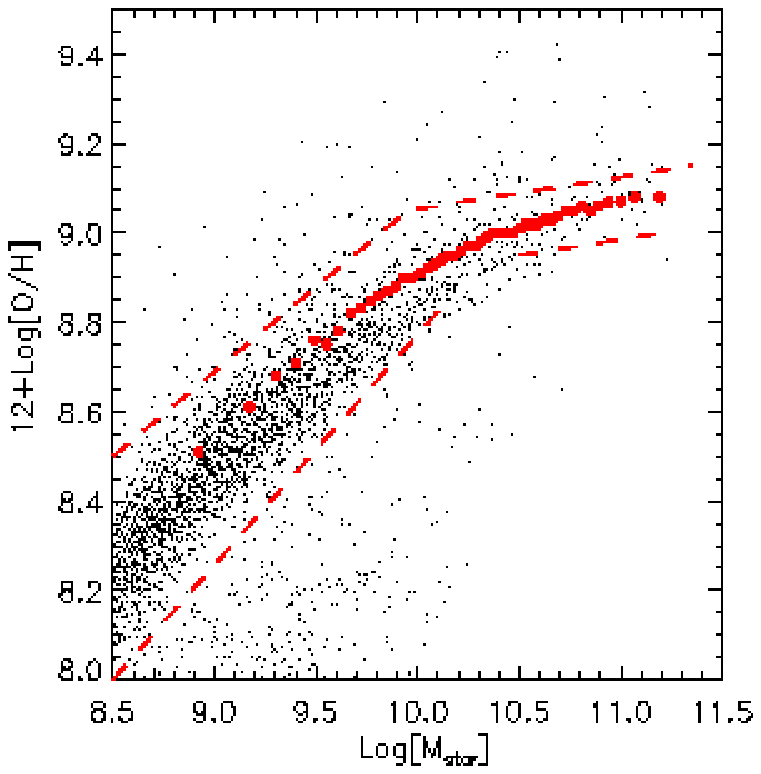}

\includegraphics{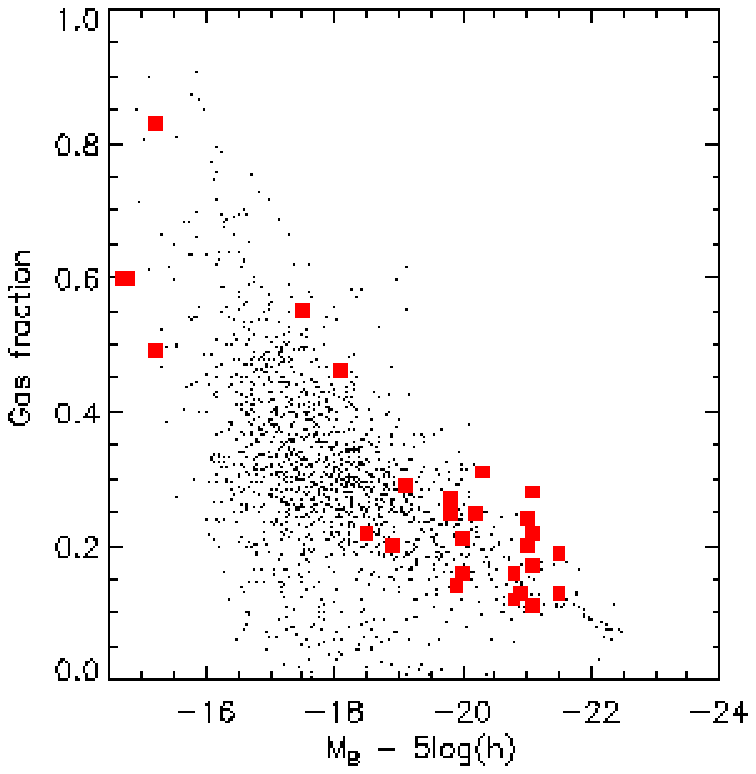}

\sidebyside
{\caption{Cold phase metallicity as a function of the stellar mass for model
    galaxies (points) compared with the results from a sample of $\sim 20000$
    galaxies from the SDSS (Tremonti et al., in preparation).}}
{\caption{Gas fraction as a function of the B--band luminosity compared with
    data from Garnett (2002).}}
\end{figure}

\section{The observations}

The ESO Distant Cluster Survey (EDisCS) is an ESO Large Programme aiming at
studying the evolution of cluster population over more that $50$ per cent of
cosmic time by combining photometric and spectroscopic information on a large
sample of clusters at redshift $\sim 0.5$ and $\sim 0.8$ with existing
information on well studied nearby clusters. 

We can compare our simulations to the data to study how the efficiency and
relative importance of processes such as quiescent and interaction-driven star
formation, ram-pressure stripping, harassment, strangulation and merging,
affect the observable structure and stellar populations of cluster galaxies.

\end{document}